\documentclass[12pt]{article}
\usepackage[english]{babel}
\usepackage{amsmath}
\usepackage{amsthm}
\usepackage{amsfonts}
\usepackage{amssymb}
\usepackage{euscript}
\usepackage[cp1251]{inputenc}
\usepackage[pdftex]{graphicx}
\newcommand{\comment}[1]{}
\usepackage[12pt]{extsizes}
\usepackage{geometry}
 \usepackage{fancyhdr}

\begin{document}
\let\thefootnote\relax\footnote{This research was supported by the Russian Science Foundation under grant 16-11-10260 and was done at the Faculty of Mechanics and Mathematics, Department of Geometry and Topology of Lomonosov Moscow State University}
\begin{center}
\textbf{\large{Matrix commuting differential operators of rank 2}}
\end{center}
\begin{center}
  \textbf{Vardan Oganesyan}
\end{center}
\begin{center}
  \textbf{Abstract}
\end{center}
In this paper we propose a very effective method for constructing matrix commuting differential operators of rank 2 and vector rank (2,2). We find new matrix commuting differential operators $L$, $M$ of orders $2$ and $2g$ respectively.\\
\begin{center}
   \textbf{Introduction}
\end{center}
Let us consider two differential operators
\begin{equation*}
L_n= \sum\limits^{n}_{i=0} u_i(x)\partial_x^i,  \quad  L_m= \sum\limits^{m}_{i=0} v_i(x)\partial_x^i,
\end{equation*}
where coefficients $u_i(x)$ and $v_i(x)$ are scalar or matrix valued functions. The commutativity condition $L_nL_m = L_mL_n$ is equivalent to a very complicated system of nonlinear differential equations. The theory of commuting ordinary differential operators was first developed in the beginning of the XX century in the works of Wallenberg \cite{Wal}, Schur ~\cite{Schur}.

If two differential operators with scalar or matrix valued coefficients commute, then there exists a nonzero polynomial $R(z,w)$ such that  $R(L_n,L_m)=0$ (see \cite{Chaundy}, \cite{Grin}). The curve $\Gamma$ defined by $R(z,w)=0$ is called the \emph{spectral curve}. If
\begin{equation*}
L_n \psi=z\psi, \quad  L_m \psi=w\psi,
\end{equation*}
then $(z,w) \in \Gamma$.

If coefficients are scalar functions, then for almost all $(z,w) \in \Gamma$, the dimension of the space of common eigenfunctions $\psi$ is the same. The dimension of the space of common eigenfunctions of two commuting scalar differential operators is called the \emph{rank} of this pair. The rank is a common divisor of $m$ and $n$. The genus of the spectral curve of a pair of commuting operators is called the genus
of this pair.

If the rank of two commuting scalar differential operators equals 1, then there are explicit formulas for coefficients of commutative operators in terms of Riemann theta-functions (see ~\cite{theta}).

The case when rank of scalar commuting operators is greater than $1$ is much more difficult. The first examples of commuting scalar differential operators of the nontrivial rank 2 and the nontrivial genus $g=1$ were constructed by Dixmier ~\cite{Dixmier} for the nonsingular elliptic spectral curve.

A general classification of commuting scalar  differential operators was obtained by Krichever \cite{ringkrichever}. The general form of commuting scalar operators of rank 2 for an arbitrary elliptic spectral curve was found by Krichever and Novikov \cite{novkrich}. The general form of scalar commuting operators of rank 3 with arbitrary elliptic spectral curve was found by Mokhov \cite{Mokhov1},\cite{Mokhov2}. In  \cite{Mironov} Mironov developed theory of self-adjoint scalar operators of rank 2 and found examples of commuting scalar operators of rank 2 and arbitrary genus. Using Mironov's method many examples of scalar commuting operators of rank 2 and arbitrary genus were found (see ~\cite{Mironov2}, \cite{Vartan}, \cite{Vartan3}, \cite{Davl}, \cite{Zegl}). Moreover, examples of commuting scalar differential operators of arbitrary genus and arbitrary rank with polynomial coefficients were constructed by Mokhov in \cite{Mokhov4}, ~\cite{Mokhov3}.

Theory of commuting differential operators helps to find solutions of nonlinear partial differential equations from mathematical physics (see ~\cite{Smirnov}, ~\cite{Smirnov4}, ~\cite{Dubrovin3}, ~\cite{Dubrovin4}). Also there are  deep connections between theory of commuting scalar differential operators and Schottky problem(see~\cite{Dubrovin2}, ~\cite{Shiota}). The theory of commuting differential operators with polynomial coefficients has connections with the Dixmier conjecture and Jacobian conjecture  (see ~\cite{belov}, ~\cite{Tsushimoto}).

A general classification of commuting matrix  differential operators  was obtained by Grinevich \cite{Grin}. Grinevich considered two differential operators
\begin{equation*}
L= \sum\limits_{i=0}^{m}U_i\partial_x^i, \quad M =\sum\limits_{i=0}^{n}V_i\partial_x^i,
\end{equation*}
where $U_i$ and $V_i$ are smooth and complex-valued $s\times s$ matrices. Let us suppose the following conditions\\
\\
1) $det(U_m)\neq 0$.\\
2) Eigenvalues $\lambda_1(x), ...,\lambda_s(x)$ of $U_m$  are distinct.   \\
3) Matrix $V_n$ is diagonalizable. Let $\mu_1(x),...,\mu_s(x)$ be eigenvalues of matrix $V_n$. Suppose that functions $\dfrac{\mu_i^m}{\lambda_i^n}$ are distinct constants  for all $i=1,...,s$. \\
\\
Easy to see that if $L$ and $M$ commute, then $U_m$ and $V_n$ commute. If operators $L$ and $M$ commute, then $FLF^{-1}$ and $FMF^{-1}$ commute, where $F$ is matrix. We also can change variable. So, without loss of generality we can suppose that
\begin{equation*}
(U_m)_{ij} =\delta_{ij}\lambda_i, \quad (V_n)_{ij} = \delta_{ij}\mu_i, \quad tr(U_{m-1})=0
\end{equation*}
Let $\Gamma$ be the spectral curve of commuting matrix operators $L, M$. Spectral curve of matrix commuting operators can be reducible. Let $\Gamma_i$ be an irreducible component of the spectral curve. The dimension of the space of common eigenfunctions
\begin{equation*}
L\psi=z\psi, \quad M\psi=w\psi, \quad (z,w)\in \Gamma_i
\end{equation*}
is called the rank of commuting pair on $\Gamma_i$. Grinevich discovered that the spectral curve $\Gamma$ has $s$ points at infinity. So, $\Gamma = \bigcup_{i=1}^k\Gamma_i$, where $k\leqslant s$. Let $l_i$ be the rank of operators on $\Gamma_i$. Operators $L, M$ are called commuting operators of vector rank $(l_1,...,l_k)$, where $k\leqslant s$. Numbers $l_i$ are common divisors of $m$ and $n$. For more details see ~\cite{Grin}. Also see ~\cite{weikard3}, ~\cite{weikard4}, ~\cite{Dubrovin1}.

If the rank of commuting matrix differential operators equals $1$, then there exists explicit formulas for coefficients in terms of Riemann theta-functions ~\cite{novkrich2}.\\

In this paper we propose a very effective method for constructing matrix commuting operators of rank 2 and vector rank (2,2). We find new commuting operators $L$, $M$ of orders $2$ and $2g$ respectively.\\

\begin{center}
\textbf{Acknowledgments}
\end{center}
The author wishes to express gratitude to Professor O. I. Mokhov for advices and help in writing this paper.

\begin{center}
  \textbf{Explicit examples of commuting matrix differential operators of rank 2}
\end{center}
Let us consider the operator
\begin{equation}
L = E(x) \partial_x^2 + R(x)\partial + Q(x),
\end{equation}
where
$$
E = \begin{pmatrix}
\lambda_1(x) & \lambda_3(x) \\
0 & \lambda_2(x)  \\
\end{pmatrix}
, \quad
R = \begin{pmatrix}
r_1(x) & r_3(x) \\
r_2(x) & -r_1(x)  \\
\end{pmatrix}
, \quad
Q = \begin{pmatrix}
q_1(x) & q_3(x) \\
q_2(x) & q_4(x)
\end{pmatrix}
$$
We want to find an operator $M$ of order $2g$ such that $[L,M]=0$. Let us consider the operator
\begin{equation}
\begin{gathered}
M = B_0(x)L^{g} + \left(A_1\left(x\right)\partial_x + B_1(x)\right)L^{g-1} + (A_2(x)\partial_x + B_2(x))L^{g-2} + ... +
\\
+(A_{g-1}(x)\partial_x + B_{g-1}(x))L + A_g(x)\partial_x + B_g(x),
\end{gathered}
\end{equation}
where
$$
A_{g-k} = \begin{pmatrix}
a_1^{g-k}(x) & a_3^{g-k}(x) \\
a_2^{g-k}(x) &a_4^{g-k}(x)  \\
\end{pmatrix}
, \quad
B_{g-k} = \begin{pmatrix}
b_1^{g-k}(x) & b_3^{g-k}(x) \\
b_2^{g-k}(x) & b_4^{g-k}(x)  \\
\end{pmatrix}
, \quad
A_{0} = \begin{pmatrix}
0 & 0 \\
0 & 0  \\
\end{pmatrix}.
$$
We note that in the formulas above the number $g-k$ is \textbf{index not degree}. The index ${g-k}$ says us that functions $a_i^{g-k}$ and $b_i^{g-k}$ are elements of matrices $A_{g-k}$ and $B_{g-k}$ respectively.

Let us note that if $[L,N]=0$, then $[L,MN]=LMN - MNL = LMN -MLN= [L,M]N$, where $L,M,N$ are matrix differential operators. So, we see that
\begin{equation*}
\begin{gathered}
\left[L,M\right] = [L,\sum\limits^{g}_{k=0}(A_{g-k}\partial_x + B_{g-k})L^k] = \sum\limits^{g}_{k=0}[L,(A_{g-k}\partial_x + B_{g-k})L^k] =\\ =\sum\limits^{g}_{k=0}[L,(A_{g-k}\partial_x + B_{g-k})]L^k.
\end{gathered}
\end{equation*}
Direct calculations show that
\begin{equation*}
\begin{gathered}
\left[L, \enskip A_{g-k}\partial_x + B_{g-k}] = [E\partial_x^2 + R\partial + Q, \enskip A_{g-k}\partial_x + B_{g-k}\right] = \\
\\
+(EA_{g-k} - A_{g-k}E)\partial_x^3 + (2EA_{g-k}' +EB_{g-k} + RA_{g-k} - A_{g-k}E' - A_{g-k}R -B_{g-k}E)\partial_x^2 + \\
+(EA_{g-k}'' + 2EB_{g-k}' + R_{g-k}A' + RB_{g-k} + QA_{g-k} - A_{g-k}R' - A_{g-k}Q - B_{g-k}R)\partial_x + \\
+(EB_{g-k}'' + RB_{g-k}' + QB_{g-k} - A_{g-k}Q' - B_{g-k}Q) = \\
\\
=K_{g-k}\partial_x^3 + P_{g-k}\partial_x^2 + T_{g-k}\partial_x + F_{g-k},
\end{gathered}
\end{equation*}
where
\begin{equation*}
\begin{gathered}
K_{g-k} = EA_{g-k} - A_{g-k}E,\\
P_{g-k} = 2EA_{g-k}' +EB_{g-k} + RA_{g-k} - A_{g-k}E' - A_{g-k}R -B_{g-k}E,\\
T_{g-k} = EA_{g-k}'' + 2EB_{g-k}' + R_{g-k}A' + RB_{g-k} + QA_{g-k} - A_{g-k}R' - A_{g-k}Q - B_{g-k}R,\\
F_{g-k} = EB_{g-k}'' + RB_{g-k}' + QB_{g-k} - A_{g-k}Q' - B_{g-k}Q.
\end{gathered}
\end{equation*}
Using the fact that $\partial_xL = E\partial_x^3 + (E' + R)\partial_x^2 + (R' + Q)\partial_x + Q'$ we get
\begin{equation*}
\begin{gathered}
K_{g-k}\partial_x^3 + P_{g-k}\partial_x^2 + T_{g-k}\partial_x + F_{g-k} = \\
\\
K_{g-k}E^{-1}\partial_xL + \left(P_{g-k} - K_{g-k}E^{-1}(E' + R)\right)\partial_x^2 + \\
\left(T_{g-k} - K_{g-k}E^{-1}(R' + Q)\right)\partial_x + (F_{g-k}  - K_{g-k}E^{-1}Q'  )=\\
\end{gathered}
\end{equation*}
\begin{equation*}
\begin{gathered}
K_{g-k}E^{-1}\partial_xL + \left(P_{g-k} - K_{g-k}E^{-1}(E' + R)\right)E^{-1}L + \\
\left(T_{g-k} - K_{g-k}E^{-1}(R' + Q) -\left(P_{g-k} - K_{g-k}E^{-1}(E' + R)\right)E^{-1}R\right)\partial_x +
+\\ \left(F_{g-l}  - K_{g-k}E^{-1}Q' - \left(P_{g-k} - K_{g-k}E^{-1}(E' + R)\right)E^{-1}Q \right) = \\
\\
\widetilde{K}_{g-k}\partial_xL + \widetilde{P}_{g-k}L + \widetilde{T}_{g-k}\partial_x + \widetilde{F}_{g-k},
\end{gathered}
\end{equation*}
where
\begin{equation}
\begin{gathered}
\widetilde{K}_{g-k} = K_{g-k}E^{-1}\\
\widetilde{P}_{g-k} = \left(P_{g-k} - K_{g-k}E^{-1}(E' + R)\right)E^{-1}\\
\widetilde{T}_{g-k} = \left(T_{g-k} - K_{g-k}E^{-1}(R' + Q) - K_{g-k}E^{-1}(E' + R)E^{-1}R\right)\\
\widetilde{F}_{g-k} =  \left(F_{g-k}  - K_{g-k}E^{-1}Q' - K_{g-k}E^{-1}(E' + R)E^{-1}Q \right).
\end{gathered}
\end{equation}
Finally we obtain
\begin{equation*}
\begin{gathered}
\left[L,M\right] = \sum\limits^{g}_{k=0}[L,(A_{g-k}\partial_x + B_{g-k})]L^k=\\
\\
=\left(\widetilde{K}_{0}\partial_x + \widetilde{P}_{0}\right)L^{g+1} + \left((\widetilde{T}_{0} + \widetilde{K}_{1} )\partial_x + (\widetilde{F}_{0} + \widetilde{P}_{1})\right)L^{g-1}+\\
+\left((\widetilde{T}_{1} + \widetilde{K}_{2} )\partial_x + (\widetilde{F}_{1} + \widetilde{P}_{2})\right)L^{g-1}+...+\\
+\left((\widetilde{T}_{g-1} + \widetilde{K}_{g} )\partial_x + (\widetilde{F}_{g-1} + \widetilde{P}_{g})\right)L + T_g\partial_x + F_g.
\end{gathered}
\end{equation*}
So,  if
\begin{equation}
\begin{gathered}
\widetilde{K}_0=0, \widetilde{P}_0=0, \widetilde{K}_1 = -\widetilde{T}_0, \widetilde{P}_1=-\widetilde{F}_0,...,\\
\widetilde{K}_m = -\widetilde{T}_{m-1}, \widetilde{P}_m = -\widetilde{F}_{m-1},...,\\
\widetilde{K}_g=-\widetilde{T}_{g-1}, \widetilde{P}_g=-\widetilde{F}_{g-1},\\
\widetilde{T}_g = 0, \widetilde{F}_g = 0,
\end{gathered}
\end{equation}
then $[L,M]=0$.

Let us calculate $\widetilde{K}_{g-k}, \widetilde{P}_{g-k}, \widetilde{T}_{g-k}, \widetilde{F}_{g-k}$ from (3). This formulas are too hard for analyzing but in the sequel only special cases are considered. We are going to show that in some cases formulas (4) give a very effective methods for finding commuting operators. Let us describe the main idea. We know that $A_0=0$ because  $M$ is operator of order $2g$. Let us take $B_0$ such that $\widetilde{K}_0 = 0$ and $\widetilde{P}_0 = 0$. Then we can calculate $\widetilde{T}_0, \widetilde{F}_0$. Using (4) we can find $\widetilde{K}_1, \widetilde{P}_1$, then $a^{1}_2, a^{1}_3, a^{1}_4, b^{1}_1, b^{1}_2, b^{1}_3, b^{1}_4$. And using (4) we can find $\widetilde{T}_1$ and $\widetilde{F}_1 $. So, we get recurrence relations
\begin{equation*}
 \begin{cases}
 a_i^{m+1} = g_i(a_1^m,a_2^m,a_3^m,a_4^m,b_1^m,b_2^m,b_3^m,b_4^m, r_1,r_2,r_3,q_1,q_2,q_3,q_4), \quad i=1,2,3,4\\
 b_i^{m+1} = h_i(a_1^m,a_2^m,a_3^m,a_4^m,b_1^m,b_2^m,b_3^m,b_4^m, r_1,r_2,r_3,q_1,q_2,q_3,q_4) \quad i=1,2,3,4.
 \end{cases}
\end{equation*}
We will see that if there exists $g$ such that
\begin{equation*}
 \begin{cases}
 a_i^{g+1} = 0\\
 b_i^{g+1} = 0
 \end{cases}
 i=1,...,4
\end{equation*}
then the operator $L$ commutes with operator $M$.\\

Let us suppose that $\lambda_1=1$, $\lambda_2 = -1$, $\lambda_3 = 0$, $r_2(x)=r_3(x)=0$, $q_3 = q_2$ and $q_4=-q_1$. Then we have
\begin{equation*}
\begin{gathered}
\widetilde{K}_{g-k} = \begin{pmatrix}
0 & -2a_3^{g-k} \\
-2a_2^{g-k} & 0
\end{pmatrix}, \quad
\widetilde{P}_{g-k} = \begin{pmatrix}
2(a_1^{g-k})' \quad & -2b_3^{g-k}   - 2(a_3^{g-k})' \\
-2b_2^{g-k}  - 2(a_2^{g-k})' \quad &   2(a_4^{g-k})'  \\
\end{pmatrix},
\\
\widetilde{T}_{g-k} = \begin{pmatrix}
 \widetilde{T}_{g-k,1}^1 \quad &  \widetilde{T}_{g-k,2}^1 \\
 \widetilde{T}_{g-k,1}^2 \quad &   \widetilde{T}_{g-k,2}^2  \\
\end{pmatrix}, \quad
\widetilde{F}_{g-k} = \begin{pmatrix}
 \widetilde{F}_{g-k,1}^1 \quad &  \widetilde{F}_{g-k,2}^1 \\
 \widetilde{F}_{g-k,1}^2 \quad &   \widetilde{F}_{g-k,2}^2\\
\end{pmatrix},
\\
\\
\widetilde{T}_{g-k,1}^1 = a_2^{g-k}q_2 + a_3^{g-k}q_2 - r_1(a_1^{g-k})' +  2(b_1^{g-k})' - a_1^{g-k}r_1' + (a_1^{g-k})'',\\
\widetilde{T}_{g-k,2}^1 = -a_1^{g-k}q_2 + a_4^{g-k}q_2 - r_1(a_3^{g-k})' +  2(b_3^{g-k})' - a_3^{g-k}r_1' + (a_3^{g-k})'',\\
\widetilde{T}_{g-k,1}^2 = a_1^{g-k}q_2 - a_4^{g-k}q_2 + r_1(a_2^{g-k})' -  2(b_2^{g-k})' + a_2^{g-k}r_1' - (a_2^{g-k})'',\\
\widetilde{T}_{g-k,2}^2 = a_2^{g-k}q_2 + a_3^{g-k}q_2 + r_1(a_4^{g-k})' -  2(b_4^{g-k})' + a_4^{g-k}r_1' - (a_4^{g-k})'',
 \end{gathered}
\end{equation*}
\begin{equation*}
\begin{gathered}
\\
\widetilde{F}_{g-k,1}^1 = b_2^{g-k}q_2 + b_3^{g-k}q_2 - 2q_1(a_1^{g-k})' + 2q_2(a_3^{g-k})' + r_1(b_1^{g-k})' - q_1'a_1^{g-k} + q_2'a_3^{g-k} + (b_1^{g-k})'',\\
\widetilde{F}_{g-k,2}^1 = -b_1^{g-k}q_2 + b_4^{g-k}q_2 - 2q_2(a_1^{g-k})' - 2q_1(a_3^{g-k})' + r_1(b_3^{g-k})' - q_1'a_3^{g-k} - q_2'a_1^{g-k} + (b_3^{g-k})'',\\
\widetilde{F}_{g-k,1}^2 = b_1^{g-k}q_2 - b_4^{g-k}q_2 + 2q_1(a_2^{g-k})' - 2q_2(a_4^{g-k})' - r_1(b_2^{g-k})' + q_1'a_2^{g-k} - q_2'a_4^{g-k} -(b_2^{g-k})'',\\
\widetilde{F}_{g-k,2}^2 = b_2^{g-k}q_2 + b_3^{g-k}q_2 + 2q_2(a_2^{g-k})' + 2q_1(a_4^{g-k})' - r_1(b_4^{g-k})' + q_1'a_4^{g-k} + q_2'a_2^{g-k} -(b_4^{g-k})''
 \end{gathered}
\end{equation*}
From (4)  we get
\begin{equation*}
\widetilde{K}_{g-k+1} = -\widetilde{T}_{g-k}, \widetilde{P}_{g-k+1} = -\widetilde{F}_{g-k}.
\end{equation*}
So, we obtain recurrence relations
\begin{equation}
\begin{gathered}
a_i^0(x) \equiv 0, \qquad i=1,...4,\\
b^0_2(x) = b^0_3 \equiv 0
\end{gathered}
\end{equation}

\begin{equation}
\begin{gathered}
\widetilde{T}_{g-k,1}^1 = 0  \Leftrightarrow  \\
b_1^{g-k} = -\dfrac{1}{2}\int\left(a_2^{g-k}q_2 + a_3^{g-k}q_2 - (a_1^{g-k})'r_1 - a_1^{g-k}r_1' + (a_1^{g-k})''\right)dx + C^{g-k}_1,
\end{gathered}
\end{equation}

\begin{equation}
\begin{gathered}
\widetilde{T}_{g-k,2}^2 = 0  \Leftrightarrow \\
b_4^{g-k} = \dfrac{1}{2}\int\left( a_2^{g-k}q_2 + a_3^{g-k}q_2 + (a_4^{g-k})'r_1 + a_4^{g-k}r_1' - (a_4^{g-k})''   \right)dx + C^{g-k}_2,
\end{gathered}
\end{equation}

\begin{equation}
\begin{gathered}
-2a_3^{g-k+1} = -\widetilde{T}_{g-k,2}^1 \Leftrightarrow \\
a_3^{g-k+1} = \dfrac{1}{2}\left(-a_1^{g-k}q_2 + a_4^{g-k}q_2 - (a_3^{g-k})'r_1 +  2(b_3^{g-k})'  -  a_3^{g-k}r_1' + (a_3^{g-k})''\right),
\end{gathered}
\end{equation}

\begin{equation}
\begin{gathered}
-2a_2^{g-k+1} = -\widetilde{T}_{g-k,1}^2 \Leftrightarrow \\
a_2^{g-k+1} = \dfrac{1}{2}\left(a_1^{g-k}q_2 - a_4^{g-k}q_2 + (a_2^{g-k})'r_1 -  2(b_2^{g-k})' + a_2^{g-k}r_1' - (a_2^{g-k})'' \right),
\end{gathered}
\end{equation}

\begin{equation}
\begin{gathered}
2(a_1^{g-k+1})' = -\widetilde{F}_{g-k,1}^1 \Leftrightarrow\\
a_1^{g-k+1} =  -\dfrac{1}{2}\int\left(b_2^{g-k}q_2 + b_3^{g-k}q_2 - 2(a_1^{g-k})'q_1 + 2q_2(a_3^{g-k})' + (b_1^{g-k})'r_1 - \right.\\ \left.
 - a_1^{g-k}q_1' +  a_3^{g-k}q_2' + (b_1^{g-k})''     \right)dx + C^{g-k+1}_3,
\end{gathered}
\end{equation}

\begin{equation}
\begin{gathered}
2(a_4^{g-k+1})' = -\widetilde{F}_{g-k,2}^2 \Leftrightarrow \\
a_4^{g-k+1} = -\dfrac{1}{2}\int\left(b_2^{g-k}q_2 + b_3^{g-k}q_2 + 2(a_2^{g-k})'q_2 + 2(a_4^{g-k})'q_1 - (b_4^{g-k})'r_1
 + \right. \\  \left. + a_4^{g-k}q_1' +   a_2^{g-k}q_2' -   (b_4^{g-k})''     \right)dx + C^{g-k+1}_4,
\end{gathered}
\end{equation}

\begin{equation}
\begin{gathered}
-2b_3^{g-k+1} - (a_3^{g-k+1})' = -\widetilde{F}_{g-k,2}^1 \Leftrightarrow \\
b_3^{g-k+1} = \frac{1}{2}\left(-b_1^{g-k}q_2 + b_4^{g-k}q_2 - 2(a_1^{g-k})'q_2 - 2(a_3^{g-k})'q_1 + (b_3^{g-k})'r_1- \right. \\ \left. - a_3^{g-k}q_1' -  a_1^{g-k}q_2' + (b_3^{g-k})''\right) - (a_3^{g-k+1})',
\end{gathered}
\end{equation}

\begin{equation}
\begin{gathered}
-2b_2^{g-k+1} - (a_2^{g-k+1})' = -\widetilde{F}_{g-k,1}^2 \Leftrightarrow \\
b_2^{g-k+1} =\dfrac{1}{2}\left( b_1^{g-k}q_2 - b_4^{g-k}q_2 + 2(a_2^{g-k})'q_1 - 2(a_4^{g-k})'q_2 - (b_2^{g-k})'r_1+ \right. \\ \left. + a_2^{g-k}q_1' - a_4^{g-k}q_2' -(b_2^{g-k})'' \right) - (a_2^{g-k+1})'.
\end{gathered}
\end{equation}
We see  that
\begin{equation*}
 \begin{cases}
   \widetilde{T}_g = 0\\
  \widetilde{F}_g = 0
 \end{cases} \quad
 \Leftrightarrow \quad
  \begin{cases}
   \widetilde{K}_{g+1} = 0\\
  \widetilde{P}_{g+1} = 0
 \end{cases}
\end{equation*}
where
\begin{equation*}
\begin{gathered}
\widetilde{K}_{g+1} = \begin{pmatrix}
0 & -2a_3^{g+1} \\
2a_2^{g+1} & 0
\end{pmatrix}, \quad
\widetilde{P}_{g+1} = \begin{pmatrix}
2(a_1^{g+1})' \quad & -2b_3^{g+1}   - 2(a_3^{g+1})' \\
-2b_2^{g+1}  - 2(a_2^{g+1})' \quad &   2(a_4^{g+1})' , \\
\end{pmatrix}
 \end{gathered}
\end{equation*}
where $C_i^j$ are arbitrary constants. Let us note that if $a_i^0(x) = 0$, $b^0_2(x)= 0,  b^0_3(x)= 0$ for all $i=1,...,4$, then from (6) and (7) we get that  $b_1^0(x) = const$ and $b_4^0(x) = const$.\\
\\
We obtain the following theorem\\

\textbf{Theorem 1.} \emph{If there exists number $g$ and constants of integration $C_j^m$ such that $a_i^{g+1}=0, b^{g+1}_2=0, b^{g+1}_3 = 0$ for all $i=1,...,4$, then  the operator  }
\begin{equation*}
L = \begin{pmatrix}
1 & 0 \\
0 & -1
\end{pmatrix}
\partial_x^2 +
\begin{pmatrix}
r_1(x) & 0 \\
0 & -r_1(x)
\end{pmatrix}
\partial_x +
\begin{pmatrix}
q_1(x) & q_2(x) \\
q_2(x) & -q_1(x)
\end{pmatrix}
\end{equation*}
\emph{commutes with operator }
\begin{equation}
M=B_0L^g + (A_1\partial_x + B_1)L^{g-1} + ... + A_0\partial_x + B_0,
\end{equation}
\begin{equation*}
\begin{gathered}
B_0=\begin{pmatrix}
\mu_1 & 0 \\
0 & \mu_2
\end{pmatrix}, \quad
B_1 = \begin{pmatrix}
C_1^1 + \dfrac{C^1_3r_1(x)}{2} & -\dfrac{(\mu_1 - \mu_2)q_2}{2} \\
\dfrac{(\mu_1 -\mu_2)q_2}{2} & C_1^4 + \dfrac{C^1_4r_1(x)}{2}
\end{pmatrix},
\end{gathered}
\end{equation*}
\emph{where $\mu_1, \mu_2$ are arbitrary constants and $C_i^j$ are some constants. We see that if $\mu_1 \neq \mu_2$ and $q_2\neq const$, then $B_1$ is not constant matrix hence $M$ is not polynomial in $L$}.\\
\\
\textbf{Theorem 2.} \emph{The operator }
\begin{equation*}
\begin{gathered}
L = \begin{pmatrix}
1 & 0 \\
0 & -1
\end{pmatrix}
\partial_x^2 +
\begin{pmatrix}
\alpha_2x^2 + \alpha_0 & 0 \\
0 & -\alpha_2x^2 - \alpha_0
\end{pmatrix}
\partial_x +
\begin{pmatrix}
\beta x^2 + \alpha_2x & \gamma x \\
\gamma x & -\beta x^2 - \alpha_2x
\end{pmatrix} ,
\end{gathered}
\end{equation*}
where
\begin{equation*}
\gamma^2 = -n^2\alpha^2_2, \quad n \in \mathbb{N}
\end{equation*}
\emph{and $\alpha_2, \alpha_0, \beta$ are arbitrary constants, commutes with differential operator (14) of order $4n$, where $g=2n$. The order of operator $M$ equals $4n$.}\\
\\
\textbf{Remark.} Calculations show that if $n \leqslant 3$, then the spectral curve of operators $L, M$ from Theorem 2 is nonsingular for almost all $\alpha_0, \alpha_2, \beta$ and is hyperelliptic. Hence $L$ and $M$ are operators of rank 2. In some cases spectral curve is reducible and we get commuting operators of rank $(2,2)$. Note that operators from Theorem 2 can't be operators of rank 1. Also note that from Theorem 1 we see that the matrix operator $M$ from Theorem 2 is operator with polynomial coefficients.
\\
\\
\textbf{Example 1.} If $n=1$ and $\mu_1=1, \mu_2=-1$, then the operator
\begin{equation*}
L = \begin{pmatrix}
1 & 0 \\
0 & -1
\end{pmatrix}
\partial_x^2 +
\begin{pmatrix}
\alpha_2x^2 + \alpha_0 & 0 \\
0 & -\alpha_2x^2 - \alpha_0
\end{pmatrix}
\partial_x +
\begin{pmatrix}
\beta x^2 + \alpha_2x & i\alpha_2 x \\
i\alpha_2 x & -\beta x^2 - \alpha_2x
\end{pmatrix}
\end{equation*}
commutes with operator $M = B_0L^2 + A_1\partial_xL + B_1L + A_2\partial_x + B_2$. Calculations show that
\begin{equation*}
\begin{gathered}
M = \begin{pmatrix}
1 & 0 \\
0 & -1
\end{pmatrix}
\partial_x^4 +
\begin{pmatrix}
2 (\alpha_2x^2 + \alpha_0)  & 0 \\
0 & -2 (\alpha_2x^2 + \alpha_0)
\end{pmatrix}
\partial_x^3 +\\
+\begin{pmatrix}
\alpha_2^2x^4 + 2(\alpha_0\alpha_2 + \beta)x^2 + 6\alpha_2x + \alpha_0^2   & i\alpha_2x \\
i\alpha_2x & -\alpha_2^2x^4 - 2(\alpha_0\alpha_2 + \beta)x^2 - 6\alpha_2x - \alpha_0^2
\end{pmatrix}
\partial_x^2 +\\
+\begin{pmatrix}
m_1  & m_2 \\
m_2 & -m_1
\end{pmatrix}
\partial_x +
\begin{pmatrix}
h_1  & h_2 \\
h_2 & -h_1 + 2\beta - \alpha_0\alpha_2
\end{pmatrix} + C_1L +
\begin{pmatrix}
C_0  & 0 \\
0 & C_0
\end{pmatrix},\\
\\
m_1 = 2\alpha_2\beta x^4 + 4\alpha_2^2x^3 + 2 \alpha_0\beta x^2  + 4(\alpha_0\alpha_2 + \beta)x +4\alpha_2,\\
m_2 = i\alpha_2^2 x^3 + i\alpha_0\alpha_2x + i\alpha_2,\\
h_1 = \beta_2^2x^4 + 4\alpha_2\beta x^3 + \dfrac{3\alpha_2^2}{2}x^2 + 2\alpha_0\beta x  + 4\beta,
\end{gathered}
\end{equation*}
\begin{equation*}
\begin{gathered}
h_2 = i\alpha_2\beta x^3 + \dfrac{3}2{}i\alpha_2^2x^2 +  \frac{i\alpha_2\alpha_0}{2},
\end{gathered}
\end{equation*}
where $C_1$ and $C_0$ are arbitrary constants.
The spectral curve of operators $L,M$ has the form
\begin{equation*}
\left(w - C_1z -( C_0 - \dfrac{\alpha_2\alpha_0 - 2\beta}{2})\right)^2 = z^4 - (\alpha_0\alpha_2 - 2\beta)z^2  - \alpha_2\alpha_0\beta + \beta^2.
\end{equation*}
If we take $C_0=\dfrac{\alpha_2\alpha_0 - 2\beta}{2}$, $C_1 =0$, then we get
\begin{equation*}
w^2 = z^4 - (\alpha_0\alpha_2 - 2\beta)z^2 - \alpha_2\alpha_0\beta + \beta^2.
\end{equation*}
This spectral curve is nonsingular if $\alpha_2\alpha_0\beta(\alpha_2\alpha_0 - \beta) \neq 0$. So in nonsingular case we get that operators $L,M$ are operators of rank 2. If $\alpha_0=0$, then the spectral curve has the form
\begin{equation}
w^2 = (z^2 + \beta)^2 \Leftrightarrow (w - z^2 - \beta)(w + z^2 + \beta)=0
\end{equation}
We see that if $\alpha_0=0$, then the spectral curve is reducible. Note that $M\neq L^2 + \beta$ and $M\neq-L^2-\beta$ but $(M - L^2 - \beta)(M + L^2 + \beta) =  0$ and we have operators of vector rank (2,2).
\\
\\
\textbf{Example 2.} If $n=1$ and $\mu_1=1, \mu_2=2$, then the operator
\begin{equation*}
L = \begin{pmatrix}
1 & 0 \\
0 & -1
\end{pmatrix}
\partial_x^2 +
\begin{pmatrix}
\alpha_2x^2 + \alpha_0 & 0 \\
0 & -\alpha_2x^2 - \alpha_0
\end{pmatrix}
\partial_x +
\begin{pmatrix}
\beta x^2 + \alpha_2x & i\alpha_2 x \\
i\alpha_2 x & -\beta x^2 - \alpha_2x
\end{pmatrix}
\end{equation*}
commutes with operator $M = B_0L^2 + A_1\partial_xL + B_1L + A_2\partial_x + B_2$. Direct calculations show that
\begin{equation*}
\begin{gathered}
M = \begin{pmatrix}
1 & 0 \\
0 & 2
\end{pmatrix}
\partial_x^4 +
\begin{pmatrix}
2(\alpha_2x^2 + \alpha_0)  & 0 \\
0 & 4(\alpha_2x^2 + \alpha_0)
\end{pmatrix}
\partial_x^3 +\\
+\begin{pmatrix}
\alpha_2^2x^4 + 2(\alpha_0\alpha_2 + \beta)x^2 + 6\alpha_2x + \alpha_0^2   & -\dfrac{i\alpha_2x}{2} \\
-\dfrac{i\alpha_2x}{2} & 2\alpha_2^2x^4 + 4(\alpha_0\alpha_2 + \beta)x^2 + 12\alpha_2x + 2\alpha_0^2
\end{pmatrix}
\partial_x^2 +\\
+\begin{pmatrix}
m_1  & m_3 \\
m_2 & m_4
\end{pmatrix}
\partial_x +
\begin{pmatrix}
h_1  & h_3 \\
h_2 & h_4
\end{pmatrix} + C_1L +
\begin{pmatrix}
C_0  & 0 \\
0 & C_0
\end{pmatrix},\\
\\
m_1 = 2\alpha_2\beta x^4 + 4\alpha_2^2x^3 + 2 \alpha_0\beta x^2  + 4(\alpha_0\alpha_2 + \beta)x +4\alpha_2,\\
m_2 = -\dfrac{i\alpha_2^2 x^3}{2} - \dfrac{i\alpha_0\alpha_2x}{2} - \frac{7i\alpha_2}{2},\\
m_3 = -\dfrac{i\alpha_2^2 x^3}{2} - \dfrac{i\alpha_0\alpha_2x}{2} + \frac{5i\alpha_2}{2},
\end{gathered}
\end{equation*}
\begin{equation*}
\begin{gathered}
m_4 = 4\alpha_2\beta x^4 + 8\alpha_2^2x^3 + 4\alpha_0\beta x^2  + 8(\alpha_0\alpha_2 + \beta)x + 8\alpha_2,\\
h_1 = \beta^2x^4 + 4\alpha_2\beta x^3 + \dfrac{3\alpha_2^2}{4}x^2 + 2\alpha_0\beta x  + \beta + \frac{3\alpha_2\alpha_0}{2},\\
h_2 = -\dfrac{i\alpha_2\beta x^3}{2}- \dfrac{9i\alpha_2^2x^2}{4} -  \frac{7i\alpha_2\alpha_0}{4},
\end{gathered}
\end{equation*}
\begin{equation*}
\begin{gathered}
h_3 = -\dfrac{i\alpha_2\beta x^3}{2} + \dfrac{3i\alpha_2^2x^2}{4} +  \frac{5i\alpha_2\alpha_0}{4}, \\
h_4 = 2\beta^2x^4 + 8\alpha_2\beta x^3 + \dfrac{9\alpha_2^2}{4}x^2 + 4\alpha_0\beta x  + 4\beta + 2\alpha_2\alpha_0,
\end{gathered}
\end{equation*}
where $C_1$ and $C_0$ are arbitrary constants. If we take $C_1=0$ and $C_0 = 0$, then the spectral curve of operators $L,M$ has the form
\begin{equation*}
16w^2 - 8w(\alpha_2\alpha_0 - 2\beta  + 6z^2) + 32z^4 + 16 (\alpha_2\alpha_0 - 2\beta)z^2 + \alpha_2^2\alpha_0^2 = 0
\end{equation*}
We see that the spectral curve is nonsingular for almost all $\alpha_2,\alpha_0.\beta$ and $L, M$ are operators of rank 2. If $\alpha_0=0$, then the spectral curve has the form
\begin{equation*}
(w-2z^2)(w-z^2 + \beta)=0
\end{equation*}
and $L, M$ are operators of vector rank $(2,2)$.
\\
\\
\textbf{Theorem 3.} \emph{Let $\wp(x)$ be the Weierstrass elliptic function satisfying the equation $(\wp'(x))^2 = 4\wp^3(x) + g_2\wp(x) $. The operator }
\begin{equation*}
\begin{gathered}
L = \begin{pmatrix}
1 & 0 \\
0 & -1
\end{pmatrix}
\partial_x^2 +
\begin{pmatrix}
0 & \alpha\wp(x) \\
\alpha\wp(x) & 0
\end{pmatrix} ,\\
\alpha^2 = 64n^4 - 4n^2, \quad n\in \mathbb{N}
\end{gathered}
\end{equation*}
\emph{commutes with a differential operator (14), of order $4n$, where $g=2n$. The order of operator $M$ equals $4n$.}
\begin{center}
  \textbf{Proofs of Theorem 2 and Theorem 3}
\end{center}

We  prove Theorem 2 and Theorem 3 using Theorem 1. Let us suppose that $C^k_2=C^k_3=C^k_4=0$ and $C_1^{2k+1}=0$ for all $k$. We know that
\begin{equation*}
\begin{gathered}
A_0 \equiv 0, \quad
B_0=\begin{pmatrix}
\mu_1 & 0 \\
0 & \mu_2
\end{pmatrix}.
\end{gathered}
\end{equation*}
Direct calculations using $(6) - (13)$ show that
\begin{equation}
\begin{gathered}
a_1^1=a_2^1=a_3^1=a_4^1=0, \\
b^1_2 = \dfrac{\mu_1 - \mu_2}{2}q_2 , \quad  b^1_3 = -\dfrac{\mu_1 - \mu_2}{2}q_2 =-b_2^1.
\end{gathered}
\end{equation}
Then
\begin{equation}
\begin{gathered}
a_1^2=a_4^2=0, \quad a^2_2=a^2_3 = -\dfrac{\mu_1 - \mu_2}{2}q_2' =   -(b^1_2)' , \\
b^2_2 = b^2_3 = \dfrac{r_1a_2^{2} - (a_2^{2})'}{2}\\
a_1^3=a_2^3=a_3^3=a_4^3=0,\\
b^3_2 = -b^3_3
\end{gathered}
\end{equation}
\textbf{Lemma 1.} If $k=2m+1$, then \\
\begin{equation}
a_1^k=a_2^k=a^k_3=a^k_4=0, \quad b_2^k =-b_3^k
\end{equation}
\emph{If $k=2m$, then}
\begin{equation}
a_1^k=a_4^k=0, \quad a^k_2=a^k_3=-(b_2^{k-1})', \quad b_3^k =b_2^k= \dfrac{r_1a_2^{k} - (a_2^{k})'}{2}
\end{equation}
\textbf{Proof}\\
We see that relations (18) and (19) is true when $k=1$ and $k=2$. Let us suppose that (18) and (19) is true for some $k$.\\
If $k=2m+1$, then using $(6) - (13)$ we get
\begin{equation*}
\begin{gathered}
a_1^{2m+2}=a_4^{2m+2}=0, \quad a^{2m+2}_2=a^{2m+2}_3=-(b_2^{2m+1})', \quad b_3^{2m+2} =b_2^{2m+2}= \dfrac{r_1a_2^{2m+2} - (a_2^{2m+2})'}{2}.
\end{gathered}
\end{equation*}
If $k=2m$, then again using $(6) - (13)$ we have
\begin{equation*}
a_1^{2m+1}=a_2^{2m+1}=a^{2m+1}_3=a^{2m+1}_4=0 \quad \quad b_3^{2m+1} = -b_2^{2m+1}.
\end{equation*}
\textbf{The Lemma is proved.}\\
\begin{flushleft}
  \textbf{Proof of Theorem 2}
\end{flushleft}
From (16) and (17) we get
\begin{equation*}
\begin{gathered}
a_1^1=a_2^1=a_3^1=a_4^1=0, \\
b^1_2 = \dfrac{\mu_1 - \mu_2}{2}q_2 = \dfrac{\mu_1 - \mu_2}{2}\gamma x, \quad  b^1_3 = -\dfrac{\mu_1 - \mu_2}{2}q_2 = -\dfrac{\mu_1 - \mu_2}{2}\gamma x.
\end{gathered}
\end{equation*}
Then
\begin{equation}
\begin{gathered}
a_1^2=a_4^2=0, \quad a^2_2=a^2_3 = -\dfrac{\mu_1 - \mu_2}{2}\gamma = -(b^1_2)' , \\
b^2_2 = b^2_3 = -\dfrac{\mu_1 - \mu_2}{4}\alpha_0\gamma   - \dfrac{\mu_1 - \mu_2}{4}\alpha_2\gamma x^2,\\
a_1^3=a_2^3=a_3^3=a_4^3=0,\\
b^3_2 = -b^3_3 =\dfrac{2C^2_1 + (\mu_1 - \mu_2)(\alpha_0\alpha_2 - 2\beta)}{4}\gamma x + \dfrac{\mu_1 - \mu_2}{4}\gamma(\gamma^2 + \alpha_2^2) x^3
\end{gathered}
\end{equation}
We  want to prove that $L$ commutes with differential operator $(14)$, where $g=2n$. From Theorem 1 and Lemma 1 we know that we must prove that there exists constants $C_1^{2k}$ such that  $b_2^{2n+1}\equiv 0$. Let us note that recurrence relations $(6) - (13)$ are linear in $a_i^{k+1}$ and $b_i^{k+1}$. Assume that $b_i^{2m-1} = x^{2m-1}$. Then we have
\begin{equation*}
\begin{gathered}
b_2^{2m} = (2m-1)(m-1)x^{2m-3} - \dfrac{\alpha_0 (2m-1)x^{2m-2}}{2} -  \dfrac{\alpha_2(2m-1)x^{2m}}{2} = -b_3^{2m},\\
a_2^{2m} = -(b_2^{2m-1})' = -(2m-1)x^{2m-2} = a_3^{2m}.
\end{gathered}
\end{equation*}
Again using $(6) - (13)$ we obtain
\begin{equation}
\begin{gathered}
a^{2m+1}_1=a^{2m+1}_2=a^{2m+1}_3=a^{2m+1}_4=0,\\
b_2^{2m+1} = \dfrac{(2m-1)(\alpha_2^2m^2 + \gamma^2)}{2m}x^{2m+1} + \\
+\dfrac{(\alpha_2\alpha_0 - 2\beta)(2m-1)^2}{2}x^{2m-1} +  \dfrac{\alpha_0^2(2m-2)(2m-1)}{4}x^{2m-3} - \\
-\dfrac{(2m-1) (2m - 2) (2m-3) (2m-4)}{4}x^{2m-5} + \dfrac{C^{2m}_1\beta}{2}x.
\end{gathered}
\end{equation}
From (20) we see that $b_2^3 = K_1^3x + K_3^3x^3$, where $K_1^3 = K_1^3(C_1^2)$ is constant and depends on $C_1^2$, $K_3^3 = \dfrac{\mu_1 - \mu_2}{4}\gamma(\gamma^2 + \alpha_2^2)$. Let us suppose that for some $m$
\begin{equation*}
b^{2m-1}_2 = K_1^{2m-1}x + K_3^{2m-1}x^3 + ... + K_{2m-1}^{2m-1}x^{2m-1},
\end{equation*}
where
\begin{equation*}
K^{2m-1}_{2m-1} = -\dfrac{(\mu_1 - \mu_2)\prod\limits_{j=1}^{m-1}\left((2j-1)(\alpha_2j^2 + \gamma^2)\right)}{2^{m}(m-1)!},
\end{equation*}
$K_{2m-3}^{2m-1}$ is constant and depends on $C_1^2$, $K_{2m-5}^{2m-1}$ is constant and depends on $C_1^2, C_1^4$, $K_{2m-2j-1}^{2m-1}$ depends on $C_1^2,...,C_1^{2j}$  and  $K_1^{2m-1}$ depends on $C_1^{2}, C_1^4,..., C_1^{2m-2}$. We see that it is true when $m=2$. Using (21) we get
\begin{equation*}
\begin{gathered}
b_2^{2m+1} = K_1^{2m+1}x + K_3^{2m+1}x^3 + ...+ K_{2m-1}^{2m+1}x^{2m-1} + K_{2m+1}^{2m+1}x^{2m+1},\\
K_1^{2m+1} = \dfrac{C^{2m}_1\beta}{2} - 30K_5^{2m-1} + \dfrac{3\alpha_0^2}{2}K_3^{2m-1} + \dfrac{\alpha_2\alpha_0 - 2\beta}{2}K_1^{2m-1}\\
K_3^{2m+1} = \dfrac{\alpha_2^2 + \gamma^2}{2}K_1^{2m-1} + \dfrac{9(\alpha_2\alpha_0 - 2\beta)}{2}K_3^{2m-1} + 5\alpha_0^2 K_5^{2m-1} - 210K_7^{2m-1}\\
...
\end{gathered}
\end{equation*}
\begin{equation*}
\begin{gathered}
K_{2m-1}^{2m+1} = \dfrac{(2m-3)(\alpha_2^2(m-1)^2 + \gamma^2)}{2m-2}K_{2m-3}^{2m-1} + \dfrac{(\alpha_2\alpha_0 - 2\beta)(2m-1)^2}{2}K_{2m-1}^{2m-1}\\
K_{2m+1}^{2m+1} = \dfrac{(2m-1)(\alpha_2^2m^2 + \gamma^2)}{2m}K_{2m-1}^{2m-1}.
\end{gathered}
\end{equation*}
Easy to see that $K_{2m-1}^{2m+1}$ depends on constant of integration $C_1^2$ because $K_{2m-3}^{2m-1}$ depends on $C_1^2$. And $K_{2m-3}^{2m+1}$ depends on $C_1^2, C_3^4$. The last coefficient $K_1^{2m+1}$ depends on constants of integrations $C_1^2,..., C_1^{2m}$.\\
Now let us consider
\begin{equation*}
b_2^{2n+1} = K_1^{2n+1}x + K_3^{2n+1}x^3 + ...+ K_{2n-1}^{2n+1}x^{2n-1} + K_{2n+1}^{2n+1}x^{2n+1}
\end{equation*}
We know that $\gamma^2 + n^2\alpha_2 = 0$ and hence $K_{2n+1}^{2n+1} = 0$. To prove Theorem 2 we must find constants $C_1^2,...,C_1^{2n}$ such that $K_1^{2n+1} = K_3^{2n+1}=...=K_{2n-1}^{2n+1}=0$. It is always possible  because $K_{2n-1}^{2n+1}$ depends n $C_1^2$ and $K_{2n-3}^{2n+1}$ depends on $C_1^2, C_3^2$ etc. The last coefficient $K_1^{2n+1}$ depends on constants of integration $C_1^2,..., C_1^{2n}$.
\begin{flushleft}
\textbf{Theorem 2 is proved.}
\end{flushleft}
\begin{flushleft}
  \textbf{Proof of Theorem 3}
\end{flushleft}
The proof of Theorem 3 coincides with the proof of Theorem 2. Let us prove that principle parts of functions $a_i^{2n+1}$ and $b_i^{2n+1}$ equals zero. We see from $(6) - (13)$ that $a_i^j$ and $b_i^j$ are elliptic functions for any $i,j$. Hence if principle parts of functions $a_i^{2n+1}$ and $b_i^{2n+1}$ equals zero, then $a_i^{2n+1}$ and $b_i^{2n+1}$ havn't poles and hence are constants. In our case these constants are zeroes.

From (16) and (17) we get
\begin{equation*}
\begin{gathered}
a_1^1=a_2^1=a_3^1=a_4^1=0, \\
b^1_2 = \dfrac{\mu_1 - \mu_2}{2}q_2 = \dfrac{\mu_1 - \mu_2}{2x^2}\alpha + O(x^2), \quad  b^1_3 = -\dfrac{\mu_1 - \mu_2}{2x^2}\alpha + O(x^2).
\end{gathered}
\end{equation*}
Then
\begin{equation}
\begin{gathered}
a_1^2=a_4^2=0, \quad a^2_2=a^2_3 = \dfrac{\mu_1 - \mu_2}{x^3}\alpha + O(x) = -(b^1_2)' , \\
b^2_2 = b^2_3 = \dfrac{3(\mu_1 - \mu_2)}{2x^4}\alpha   + \dfrac{(\mu_1 - \mu_2)g_2\alpha}{40} + O(x^4),\\
a_1^3=a_2^3=a_3^3=a_4^3=0,\\
b^3_2 = -b^3_3 =\dfrac{(\mu_1 - \mu_2)\alpha(\alpha^2 - 60)}{4x^6}  + \dfrac{\alpha(80C_1^2 + 6(\mu_1 - \mu_2)g_2\alpha^2)}{160x^2} + O(x^2)
\end{gathered}
\end{equation}
We mentioned before that  recurrence relations $(6) - (13)$ are linear in $a_i^{k+1}$ and $b_i^{k+1}$. Assume that $b_3^{2m-1} = -b_2^{2m-1}= \dfrac{1}{x^{4m - 2}}$ and $a_1^{2m-1}=a_2^{2m-1}=a_3^{2m-1}=a_4^{2m-1}=0$. Then we have
\begin{equation*}
\begin{gathered}
b_2^{2m}=b_3^{2m} = \dfrac{8m^2 - 6m + 1}{x^{4m}},\\
a_2^{2m} = -(b_2^{2m-1})' = \dfrac{4m-2}{x^{4m-1}}.
\end{gathered}
\end{equation*}
Again using $(6) - (13)$ we obtain
\begin{equation}
\begin{gathered}
a^{2m+1}_1=a^{2m+1}_2=a^{2m+1}_3=a^{2m+1}_4=0,\\
b_2^{2m+1} = \dfrac{(2m-1)(\alpha^2 + 4m^2 - 64m^4)}{2mx^{4m+2}} + \dfrac{const}{x^{4m-2}} + \dfrac{const}{x^{4m-6}}+...+\dfrac{C_1^{2m}\alpha}{2x^2} + O(x^2)
\end{gathered}
\end{equation}
From (22) we see that $b_2^3 = \dfrac{K_6^3}{x^6} + \dfrac{K_2^3}{x^2} + O(x^2)$, where $K_2^3 = K_2^3(C_1^2)$ is constant and depends on $C_1^2$, $K_6^3 = \dfrac{(\mu_1 - \mu_2)\alpha(\alpha^2 - 60)}{4}$. Let us suppose that for some $m$
\begin{equation*}
b^{2m-1}_2 = \dfrac{K_{4m-2}^{2m-1}}{x^{4m-2}} + \dfrac{K_{4m-6}^{2m-1}}{x^{4m-6}} + ... + \dfrac{K_{2}^{2m-1}}{x^2} + O(x^2),
\end{equation*}
where
\begin{equation*}
K^{2m-1}_{4m-2} = -\dfrac{(\mu_1 - \mu_2)\prod\limits_{j=1}^{m-1}\left((2j-1)(\alpha^2 + 4j^2 - 64j^4))\right)}{2^{m}(m-1)!},
\end{equation*}
$K_{4m-6}^{2m-1}$ is constant and depends on $C_1^2$, $K_{4m-10}^{2m-1}$ is constant and depends on $C_1^2, C_1^4$, $K^{2m-1}_{4m-4j-2}$ depends on $C_1^2,...,C_1^{2j}$ and  $K_2^{2m-1}$ depends on $C_1^{2}, C_1^4,...,C_1^{2m-2}$. We see that it is true when $m=2$. Using (23) we get
\begin{equation*}
b_2^{2m+1} = K_1^{2m+1}x + K_3^{2m+1}x^3 + ...+ K_{2m-1}^{2m+1}x^{2m-1} + K_{2m+1}^{2m+1}x^{2m+1}.
\end{equation*}
Easy to see that $K_{4m-2}^{2m+1}$ depends on constant of integration $C_1^2$. And $K_{4m-6}^{2m+1}$ depends on $C_1^2, C_3^2$. The last coefficient $K_2^{2m+1}$ depends on constants of integrations $C_1^2,..., C_1^{2m}$.\\
Now let us consider
\begin{equation*}
b_2^{2n+1} = K_1^{2n+1}x + K_3^{2n+1}x^3 + ...+ K_{2n-1}^{2n+1}x^{2n-1} + K_{2n+1}^{2n+1}x^{2n+1}
\end{equation*}
We know that $\alpha^2 + 4n^2 - 64n^4 = 0$  hence $K_{2n+1}^{4n+2} = 0$. To prove Theorem 3 we must find constants $C_1^2,...,C_1^{2n}$ such that $K_2^{2n+1} = K_6^{2n+1}=...=K_{4n-2}^{2n+1}=0$. It is always possible because $K_{4n-2}^{2n+1}$ depends on $C_1^2$ and $K_{4n-2}^{2n+1}$ depends on $C_1^2, C_3^2$ etc. The last coefficient $K_2^{2n+1}$ depends on constants of integration $C_1^2,..., C_1^{2n}$.
\begin{flushleft}
\textbf{Theorem 3 is proved.}
\end{flushleft}

Department of Geometry and Topology, Faculty of Mechanics and Mathematics, Lomonosov Moscow State University, Moscow, 119991 Russia.\\
\\
E-mail address: vardan.o@mail.ru

\end{document}